\documentclass[aps,twocolumn,showrefs,amsmath,amssymb,pre,colors]{revtex4-1}

\usepackage{amssymb,amsmath,amsfonts}  
\usepackage{graphicx,color}
\usepackage{hyperref}
\usepackage[english]{babel}
\usepackage{tabularx}
\usepackage{bm}
\usepackage{calrsfs}
\usepackage{csquotes}
\def\rb{\mathbf{r}}

\def\be{\begin{equation}}
\def\ee{\end{equation}}
\def\ba{\begin{eqnarray}}
\def\ea{\end{eqnarray}}
\def\<{\left<}
\def\>{\right>} 
\def\({\left(}
\def\){\right)} 
\def\[{\left[}
\def\]{\right]} 

\def\Dg{\Delta}

\begin{document}

\title{Thermoelectric ratchet effect for charge carriers with hopping dynamics}

\author{Alois W\"urger} 
\affiliation{Universit\'e de Bordeaux $\&$ CNRS, LOMA (UMR 5798), 33405 Talence, France}

\begin{abstract}
We show that the huge Seebeck coefficients observed recently for ionic conductors, arise from a ratchet effect where activated jumps between neighbor sites are rectified by a temperature gradient, thus driving mobile ions towards the cold. For complex systems with mobile molecules like water or polyethylen glycol, there is an even more efficient diffusiophoretic transport mechanism, proportional to the thermally induced concentration gradient of the molecular component. Without free parameters, our model describes experiments on the ionic liquid EMIM-TFSI and hydrated NaPSS, and it qualitatively accounts for polymer electrolyte membranes with Seebeck coefficients of hundreds of $k_B/e$.  \end{abstract}  

\maketitle

{\it Introduction.} -- Thermoelectric materials are widely investigated in view of applications for harvesting waste heat, temperature sensors,  and cooling devices \cite{Bell2008}, where the underlying conductivity mechanism may be electronic \cite{Bell2008,Bubnova2011,Russ2016,Gregory2018,Xiao2017}, ionic \cite{Zhao2016,Kim2016,Wang2017,Li2019,Zhao2019,Han2020}, or both \cite{Wang2015,Chang2016}.  As their fundamental property, an applied temperature difference induces an electric current or, in the ionic case, a thermoelectric field. According to the second law, heat flows towards lower temperatures; thus positive and negative charge carriers are dragged by their ``heat of transport''  $Q_\pm$  \cite{deGroot1962}, generating a thermopower which is quantified by the Seebeck coefficient $S=(Q_+ - Q_-)/2eT$. 
 
Most metals and conventional semiconductors show a weak thermoelectric response, as expected for a band structure with nearly symmetric electron and hole states \cite{Fritsche1971}. Their thermopower contributions $Q_\pm$ largely cancel, resulting in $S\ll k_B/e=86\mathrm{\mu V/K}$. Higher values $S\sim10k_B/e$ are observed for doped organic polymers \cite{Bubnova2011,Russ2016,Gregory2018}, where  structural disorder and electron-phonon coupling concur to hopping conductivity, with the heat of transport given by the dissociation energy \cite{Mott1971}.

In recent years Seebeck coefficients of up to  $300 k_B/e$ were reported for ionic conductors such as hydrated polystyrene sulfonate \cite{Kim2016,Wang2017}, cellulosic membranes infiltrated with electrolyte solution \cite{Li2019}, ionic liquids \cite{Zhao2019} and polymer gels \cite{Zhao2016,Han2020}.  The mechanisms leading to these huge Seebeck coefficients are poorly understood. Thermodiffusion in dilute electrolytes, as laid out by Nernst \cite{Nernst1889}, is not valid for such complex materials and does not account for the large values of $S$: Ionic heats of transport $Q_\pm$ in aqueous solutions are of order of the thermal energy $k_BT$ \cite{Agar1989}, resulting in $S\sim k_B/e$, as confirmed by thermoelectric effects on colloidal suspensions \cite{Putnam2005,Wuerger2010,Vigolo2010,Majee2011,Eslahian2014}.  

Here we propose a model for the thermoelectric properties for charge carriers with hopping dynamics. We identify a ratchet mechanism which rectifies thermally activated jumps between neighbor sites and which relates the Seebeck coefficient to the enthalpy barrier. It turns out that companion fields of the temperature gradient, for example the thermally induced concentration gradient of a molecular component, may strongly enhance the thermoelectric response. We give a detailed comparison of our findings with experimental data for the above mentioned materials. We discuss a simple physical picture which could be relevant for both ionic and electronic polymer-based conductors. 
\begin{figure}[b]
\includegraphics[width=\columnwidth]{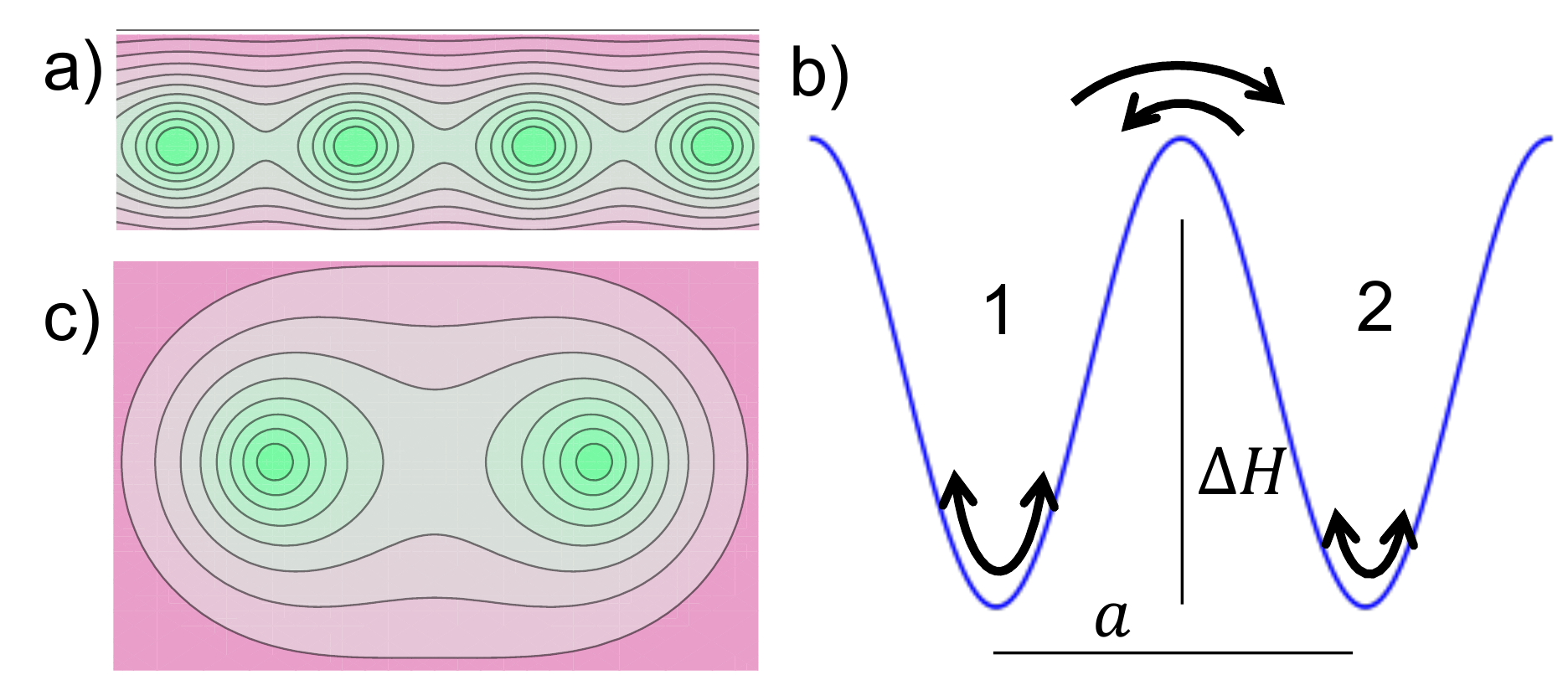}
\caption{a) Gibbs enthalpy landscape of a mobile charge in a conduction channel. The carrier passes from one site to the other by thermally activated jumps over the barrier. b) Non-equilibrium hopping dynamics between neighbor sites. The temperature in the left well is higher than in the right one, $T_1>T_2$, resulting in stronger thermal fluctuations and more frequent jumps to the right, $\Gamma_{12}>\Gamma_{21}$. c) Gibbs enthalpy landscape of an isolated double-well potential.} 
\label{figure1}
\end{figure}

{\it Hopping dynamics and Seebeck coefficient.} -- We consider carriers of charge $q$ which are trapped at well-defined sites but perform thermally activated jumps to neighbor sites.  According to the Eyring model \cite{Eyring1935}, the jump rate reads as 
   \be
   \Gamma = \Gamma_0 e^{-\Dg H/k_BT} e^{\Dg S/k_B} , 
  \label{eq:2} 
 \ee
where the attempt frequency $\Gamma_0= k_BT/h$ is  the ratio of the thermal energy and Planck's constant. The Gibbs energy barrier $\Dg G=\Dg H - T\Dg S$ consists of the activation enthalpy $\Dg H$ and entropy $\Dg S$; the former is given by the height of the barrier illustrated in Fig. 1, and the latter is related to the width of the valley at the saddle point. This picture is supported by conductivity measurements on amorphous polyelectrolyte complexes \cite{De2017,Ostendorf2019}, which provide evidence that charge transport occurs through activated jumps of the cations between discrete sites. 

The exponential factor in (\ref{eq:2}) gives the probability of a local free enthalpy fluctuation which eventually permits a mobile charge to cross the barrier. At thermal equilibrium, forward and backward jumps between two degenerate wells occur at the same rate,  $\Gamma_{12}=\Gamma_{21}$. In a non-uniform temperature as illustrated in Fig. 1b, however, jumps from a given site $1$ to the neighbor site $2$ occur at rate $\Gamma_{12}=\Gamma(T_1)$ that depends on the temperature $T_1$ of the initial position, whereas the inverse rate reads $\Gamma_{21}=\Gamma(T_2)$. The difference arises mainly from the variation of the Planck-Plesset or entropy potential $\Dg G/T$. Because of the very small temperature difference $T_1-T_2=\mathbf{a}\cdot\nabla T<1$ mK, the change of the exponent in (\ref{eq:2}) is much smaller than unity \cite{SI-TE2020}. Then we may safely linearize the rate difference,   
\be
 \frac{ \Gamma_{12}(T_1) - \Gamma_{21}(T_2)}{\Gamma} 
         =  \(1 +  \frac{\Dg H}{k_BT} \) \frac{T_1 - T_{2}}{T},
\label{eq:4}
\ee
with mean values $\Gamma$ and $T$. (For the sake of simplicity we consider degenerate wells; yet this result is readily generalized to the case of a finite bias energy \cite{SI-TE2020}.) Thus the temperature gradient rectifies the thermal charge fluctuations due to activated hopping, as discussed previously for impurity atoms in crystals \cite{LeClaire1954,Brinkman1954} 
 
\begin{figure}[t] 
\includegraphics[width=\columnwidth]{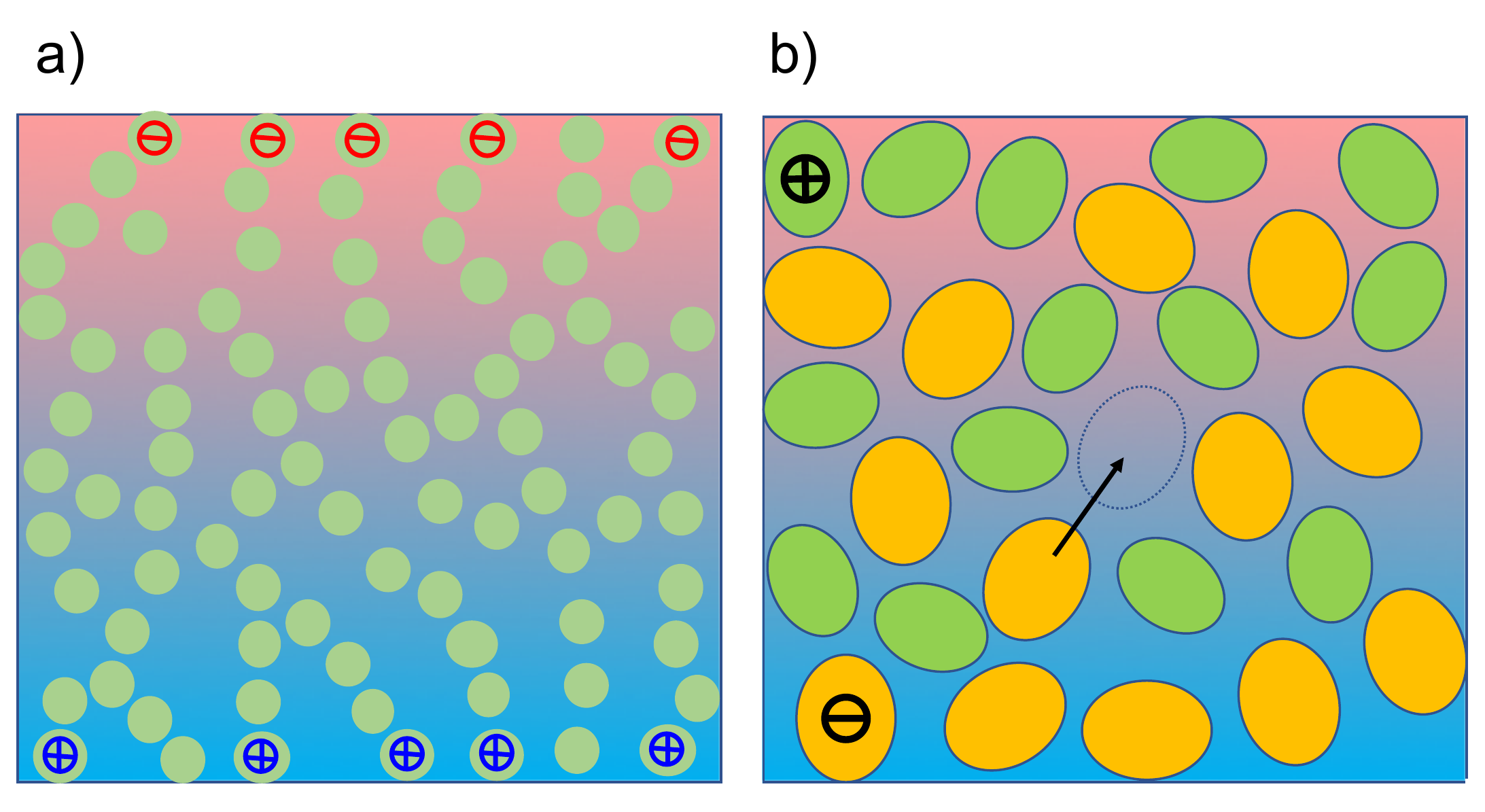}
\caption{a) Thermodiffusion in a connected network like in Fig. 1a, for example of sodium ions in NaPSS \cite{De2017}. The cations diffuse towards the cold boundary and leave uncompensated PSS$^-$ at the hot side, thus inducing a thermoelectric field $E=S\nabla T$. b)  In ionic liquids, charge transport relies on vacancy formation \cite{Abbott2004}; initial and final states are shown schematically in Fig. 1c. We show the case realized in EMIM$^+$-TFSI$^-$, where anions (yellow) accumulate at the cold side, and cations (green) at the hot.}
\label{default}
\end{figure}

For a macroscopic network of connected paths as shown in Fig. 2a, the rate difference of backward and forward jumps results in a drift velocity $v_T=-a(\Gamma_{12}-\Gamma_{21})$. Inserting (\ref{eq:4}) we find $v_T = - \mu Q\nabla T/T$, with the mobility $\mu$ and the heat of transport
  \be
  Q_\pm = k_BT + \Dg H_\pm, 
  \label{eq:7b}
  \ee  
which is positive for charges $q$ of either sign. Charge transport in ionic liquids is described by the thermally activated formation of short-lived vacancies \cite{Abbott2004}, as illustrated in Fig. 2b. Averaging over the vacancy position, we recover the above description of a connected network, albeit with $\Delta H_\pm$ comprising the (positive) barrier for ionic motion and the (negative) vacancy enthalpy. For impurity atoms in crystals, these contributions to $\Delta H$ have been determined experimentally \cite{Jaffe1964}.

According to the second law, the excess enthalpy $Q_\pm$ has a tendency to flow towards the cold, thus dragging the ions opposite to the temperature gradient. This thermodiffusion current results in  surface charges at the hot and cold sample boundaries (Fig 2a), which in turn give rise to an electric field $E$ and opposite velocity $v_E= q E$. In a closed system, the steady state is achieved if the total ion velocity vanishes, $\mu(q E-Q_\pm\nabla T/T)=0$, implying generation of a thermoelectric field $E=S\nabla T$, with the Seebeck coefficient $S =  Q_\pm/qT$.  

In the presence of monovalent charges of either sign $\pm e$, the Seebeck coefficient 
  \be
  S = \frac{Q}{eT}
  \label{eq:7}
  \ee  
is given by the mean heat of transport
  \be
  Q = w_+Q_+ - w_- Q_- ,
  \label{eq:7c}
  \ee  
weighted with the fraction of ions $w_\pm=n_\pm/(n_++n_-)$ contributing to thermally driven charge transport \cite{Majee2011}. Note that the Hittorf transport numbers for the conductivity, $t_\pm=\mu_\pm n_\pm/(\mu_+n_+ + \mu_- n_-)$, include the mobility \cite{Han2020a}.

\begin{table}[b]
\caption{Comparison of the heat of transport $Q$ obtained from Seebeck data according to (\ref{eq:7}), and the measured activation enthalpy obtained from the conductivity or, in the case of EMIM-TFSI, from the ion mobilities $\mu_\pm$.   }
\def\IL{1-ethyl-3-methylimidazolium [EMIM] -bis(trifluoro-methylsulfonyl)imide[TFSI]}
\def\cp{Poly(vinylidene fluoride-co-hexafluoropropylene)}
\def\PEG{Polyethylene-glycol 300 Da}
\def\ft1{Measured for (NaPSS)$_{0.55}$PDADMAC$_{0.45}$ \cite{De2017}, see Fig. 3.}
\begin{tabular}{|l|c|c|c|}
\hline
                    &  $S$ (mV/K)          &\; $Q (\mathrm{eV})$\; & $\Dg H (\mathrm{eV})$  \\   \hline    
    EMIM-TFSI\footnote{\IL} 
                      & $-0.85$ \cite{Zhao2019} &  $-0.255$  & 0.27/0.26 \cite{Agostino2018} \\ \hline 
    EMIM-TFSI-cp\footnote{\cp}-PEG\footnote{\PEG}   
                                                   & $-4...12$ \cite{Zhao2019}    & $-1.2 ... 3.6$ &     \\ \hline        
    NaPSS-water    & 4 \cite{Wang2017}  &   1.2           & 0.5\footnote{\ft1} \cite{De2017}    \\ \hline
   KCl-gelatin  &  6.7 \cite{Han2020}              & 2    &                                       \\   \hline    
    NaPEG-OH & 11 \cite{Zhao2016,Li2019} &    3.3       &              0.5 \cite{Zhao2016}    \\ \hline
   NaPEG-OH-cellulose & 24 \cite{Li2019} &    7.2       &                         \\ \hline
 \end{tabular}
\linebreak
\end{table}

Eqs. (\ref{eq:7b}) and (\ref{eq:7}) relate the Seebeck coefficient to the enthalpy barrier.  The latter is in general much larger than the thermal energy, resulting in $S\gg k_B/e$ and thus providing a rationale for the large Seebeck coefficients of ionic conductors listed in Table 1. For the pure ionic liquid EMIM$^+$-TFSI$^-$, the measured heat of transport $Q_\mathrm{exp}=-0.255$ eV is very close to the activation enthalpies of cations and anions \cite{Agostino2018}. A quantitative comparison is obtained from Eq. (\ref{eq:7c}); inserting the barriers $\Dg H_\pm$ and weights $w_-=0.93$, deduced from $t_-=0.9$ \cite{Tafur2015} and the mobility ratio $\mu_+/\mu_-=1.3$ \cite{Zhao2019}, we obtain the theoretical value  $Q_\mathrm{th} = - 0.25$ eV. This very good agreement provides evidence for the ratchet mechanism (\ref{eq:4}) being at the origin of the large Seebeck coefficient $S\approx-10 k_B/e$. On the other hand, in view of (\ref{eq:7c}), the relation $Q\approx-\Dg H_-$ implies $w_-\gg w_+$, thus confirming that anions are the majority carriers. 

The remaining systems of Table 1 are more complex, consisting of mobile ions and polymer matrices with infiltrated water or PEG. They do not satisfy the relation (\ref{eq:7b}): Though the ions move through activated jumps, the heat of transport by far exceeds the enthalpy barrier. In the following we show how the added nonionic molecular component modifies the thermodynamic response.

{\it Non-equilibrium composition.} -- Polyelectrolyte materials may absorb a significant weight fraction $c$ of water \cite{De2017}, which modifies both thermoelectric \cite{Kim2016,Wang2017} and transport properties \cite{Michaels1965,Varcoe2007,De2017,Ostendorf2019}. Similarly, complex polymer electrolytes are very sensitive to the composition: When adding PEG, the Seebeck coefficient of (EMIM-TFSI)-(PVDF-HFP) changes sign and increases from $-4$  to $+12$ mV/K \cite{Zhao2019}; for KCl-gelatin, $S$ shows a strong and nonlinear variation with the salt concentration \cite{Han2020}. These observations cannot be explained in terms of (\ref{eq:7b}) with a composition dependent $\Dg H(c)$: The measured heat of transport would imply enthalpy barriers of several eV (Table 1), which would inhibit ion diffusion and is clearly not compatible with the measured conductivity. 

Yet the above description based on the rate difference (\ref{eq:4}) is valid only if the composition $c$ is not affected by the non-uniform temperature.  In the general case, when evaluating the transport properties in a temperature gradient, we need to account for thermodiffusion of molecular components such as water or PEG, which gives rise to a non-uniform concentration $c(T(\rb))$. Then the activation enthalpy and entropy $\Dg H(c)$ and $\Dg S(c)$ are not the same in two neighbor wells, and the rate difference of backward and forward jumps reads 
\be
 \frac{ \Gamma_{12} - \Gamma_{21}}{\Gamma} 
         =    \frac{Q}{k_BT}  \frac{T_1-T_2}{T} - \frac{G_c}{k_BT} (c_1-c_2)  ,
\label{eq:14}  
\ee 
with $G_c=H_c-TS_c$ and the derivates 
   \be
   H_c = \frac{d\Dg H}{dc}, \;\;\; S_c = \frac{d\Dg S}{dc} .
   \label{eq:11}
   \ee

Defining the gradient $c_1-c_2=\mathbf{a}\cdot\nabla c$, we find that the ion acquires a drift velocity  $v_c=\mu G_c\nabla c$, in addition to $v_T$ defined below (\ref{eq:4}). Proceeding as above Eq. (\ref{eq:7}), we obtain the effective Seebeck coefficient 
\be
  S  = \frac{Q_\mathrm{eff}}{qT}
\label{eq:16}
\ee
with 
\be
  Q_\mathrm{eff}  =  Q - (H_c-TS_c) Tc_T  .
\label{eq:17}
\ee
and the temperature derivativee 
   \be
   c_T = \frac{dc}{dT} .
   \label{eq:19}
   \ee
Here $Q(c)=\Dg H(c)+k_BT$ is the actual heat of transport as measured in terms of the Peltier coeffiicient $\Pi=Q/e$ \cite{deGroot1962}, whereas the additional term in (\ref{eq:17}) accounts for the modified hopping dynamics due to the non-uniform composition.

\begin{figure}[t]
\includegraphics[width=\columnwidth]{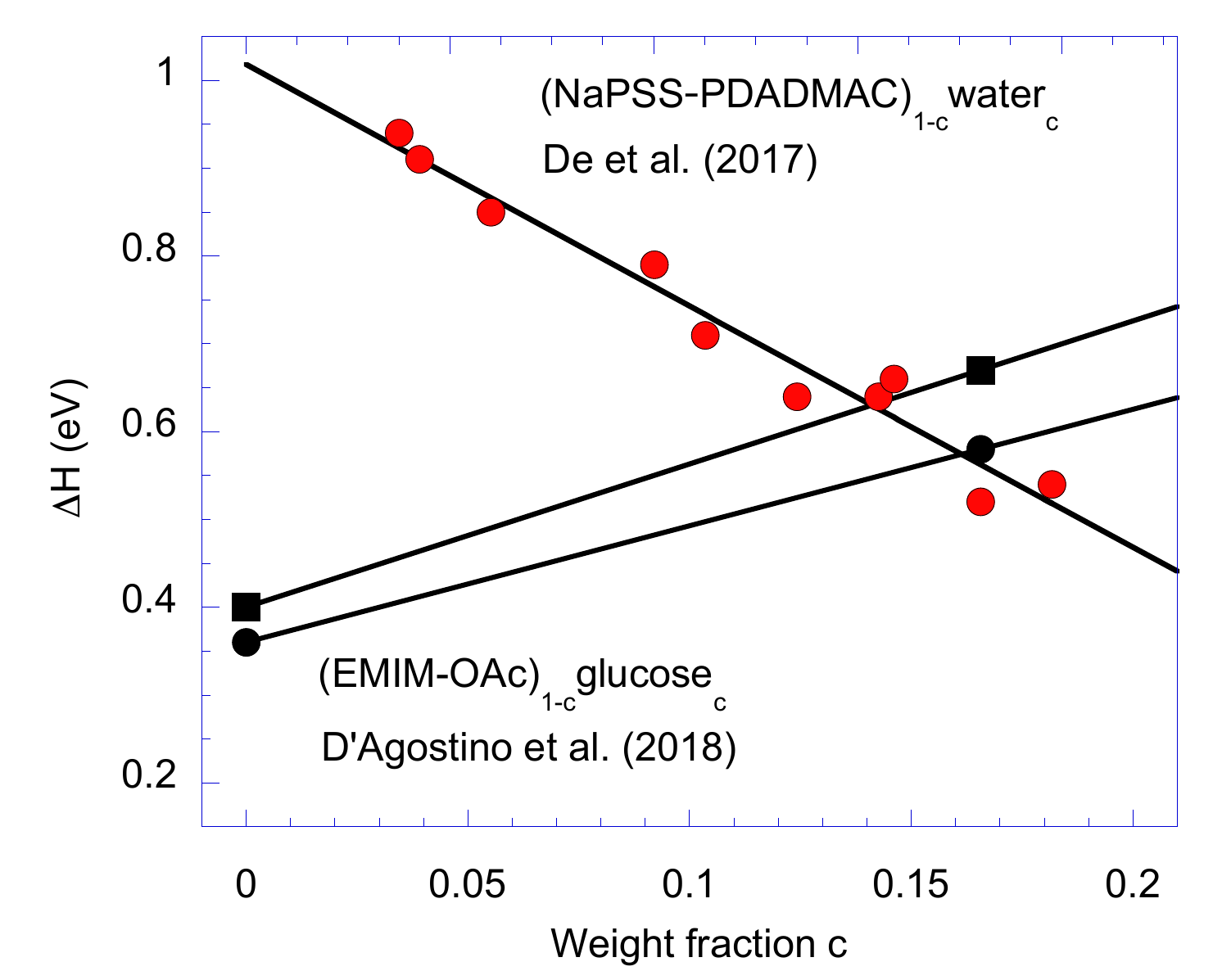}
\caption{Composition dependence of the activation enthalpy of mobile ions. Red circles show data of De et al. \cite{De2017} for sodium ions in the polyelectrolyte complex NaPSS$_{0.55}$-PDADMAC$_{0.45}$ with water weight fraction $c$. Black circles and squares give data for EMIM$^+$ and acetate$^-$, with or without added glucose, as measured by D'Agostino et al. \cite{Agostino2018}. The solid lines indicate linear fits, with slope $H_c=-2.75$ eV (Na), 1.35 eV (EMIM), and 1.66 eV (acetate).}
\label{figure3}
\end{figure} 

In order to obtain an estimate for the relevance of the diffusiophoretic contribution to the heat of transport (\ref{eq:17}), we plot in Fig. 3 the composition dependence of the enthalpy barrier for a hydrated polyelectrolyte complex and the ionic liquid EMIM-OAc with added sugar. The slope of the solid lines gives the derivative $H_c$, which varies for these ions from $-2.75$ to $+1.66$ eV. There seem to be no data for the derivative $c_T$. For binary liquids this parameter is related to the well-known Soret coefficient $S_T=c_T/(c(1-c))$; with typical values $S_T\sim 10^{-2}$ K$^{-1}$ \cite{Wiegand2004}, we find $Tc_T\sim1$, where either sign is possible. This estimate suggests that the composition-driven term in (\ref{eq:17}) may attain several eV. 

As an example we consider hydrated NaPSS, with a maximum Seebeck coefficient of 4 mV/K \cite{Wang2017}. The analysis in Ref. \cite{De2017} suggests a water content of $c\approx0.2$; as an estimate we take the enthalpy barrier of NaPSS-PDADMAC in Fig. 3, $\Dg H\approx0.5$ eV. The difference to the measured heat of transport $Q_\mathrm{exp} =1.2$ eV, is met by the diffusiophoretic contribution with the above parameters,  $- H_c Tc_T\sim0.7$ eV. 

In the first part of this paper we have shown that thermodiffusion of ions with activated dynamics is much stronger than in ordinary liquids. Yet this result holds true also for non-ionic molecules such as water, PEG, and glucose -- the enthalpy barrier of glucose in EMIM-OAc is 0.38 eV \cite{Agostino2018} --, and should result in larger values for $Tc_T$ than the above estimate. This would explain the very large Seebeck coefficients measured for the polymer electrolyte Na$^+$PEG-OH$^-$ in cellulose matrices \cite{Li2019}. In physical terms, it would imply that ion thermodiffusion in these membranes is mainly due to diffusiophoresis, at a velocity $v_c=\mu H_c \nabla c$, which is driven by the thermally induced concentration gradient of water, $\nabla c = c_T\nabla T$.

{\it Discussion.} -- The ratchet effect expressed by Eq. (\ref{eq:4}) and illustrated in Fig. 1, relies on two assumptions: First, the ions move through jumps between discrete positions and, second, forward and backward hopping rates $\Gamma_{12}(T_1)$ and $\Gamma_{21}(T_2)$ depend on the temperature of the initial well.  

The first condition requires that the activated behavior is related to jumps between discrete sites, as opposed to thermodiffusion in aqueous solution, where the ion's motion is phoretic, creating a steady backward flow in a homogeneous medium \cite{Morthomas2008}. Most systems of Table 1 are based on solid polymer matrices; adding small amounts of water, PEG, or ionic liquid, is not likely to create fluid domains, suggesting that charge carriers move through discrete jumps. Recent attempts to rationalize the structural and dynamical properties of ionic liquids like EMIM-TFSI \cite{Abbott2004}, rely on  F\"urth's hole theory for liquids \cite{Fuerth1941}, where conductivity arises from the formation of short-lived vacancies illustrated in Fig. 2b.  

The second condition is rather obviously satisfied in Eyring's transition state theory \cite{Eyring1935}, where barrier crossing is a mechanical problem and where the required energy is provided in the initial state by collisions with other atoms or, in a solid, with collective modes. For the case of strongly damped motion, Kramers \cite{Kramers1940} showed that barrier crossing results from Brownian motion, and also occurs with an activated rate. The required energy is taken from thermal Langevin forces along the uphill part of the trajectory, resulting in an asymmetry of forward and backward rates similar to (\ref{eq:4}). 

The discrete model studied here, provides a simple physical picture for the ratchet effect and the  strong thermopower observed in ionic conductors: Each jump of the mobile ion collects energy of the  order of the barrier height $\Dg H$ at the initial position and releases the same amount in the neighbor well. According to Clausius' formulation of the second law, heat necessarily flows from the hot to the cold, implying that jumps opposite to the temperature gradient are favored, and that the rate imbalance is proportional to the enthalpy barrier. As a consequence, the heat of transport is given by $\Dg H$. 

For complex materials containing mobile molecular components, we find in addition to the thermal ratchet effect, a diffusiophoretic mechanism of ion transport: The applied temperature gradient induces a non-uniform composition $c$, for example, a concentration gradient of hydration water or polyethylene glycol. Then the free enthalpy barrier $\Dg G(c)$ is different for forward and backward jumps, and the effective heat of transport $Q_\mathrm{eff}$ includes a term proportional to the concentration derivative of the free enthalpy of activation $G_c$. Formally the correction term in (\ref{eq:17}) can be expressed as the difference of the enthalpy at variable and constant $c$,
  \be
  Q_\mathrm{eff} - Q = - \left. T^2\frac{d}{dT} \frac{G}{T} \right|_{c=c(T)} 
                                +   \left. T^2\frac{d}{dT}\frac{G}{T}  \right|_{c=\mathrm{const.}}
  \label{eq:21}
  \ee
The second term on the right-hand side is the usual Gibbs-Helmholtz expression for the equilibrium enthalpy, whereas the first one allows for a non-uniform concentration, which gives rise to the drift velocity $v_c=\mu G_c \nabla c$, with $\nabla c = c_T\nabla T$. Similar effects are known for colloidal thermophoresis in liquids, where particle motion is often dominated by concentration gradients of added polymer \cite{Jiang2009}, salt \cite{Eslahian2014}, or a binary liquid at the critical point \cite{Buttinoni2012,Wuerger2015}. 

Regarding electronic systems, the binding energy of small polarons in doped organic polymers \cite{Gregory2018} and metal oxides \cite{Xiao2017} contributes to $\Dg H$ but is absent in the heat of transport $Q$.  On the other hand, the electronic heat of transport in pure silicon is dominated by the phonon-drag contribution, which accounts for electrons being nudged by long-lived thermal phonons \cite{Trzcinski1986}. 

Summarizing our comparison with experiments on ionic conductors, we obtain a quantitative description for the thermopower $S$ of the pure ionic liquid EMIM-TFSI, without free parameter, by identifying the heat of transport with the enthalpy of activation. Regarding the hydrated polyelectrolyte NaPSS, our results show that thermally induced diffusiophoresis largely contributes to the effective heat of transport (\ref{eq:17}) and thus to the Seebeck coefficient. An estimate of the parameters of Eq. (\ref{eq:17}), suggests that ion diffusiophoresis due to the gradient in water content, accounts for the large Seebeck coefficients measured for electrolyte infiltrated membranes.  

Helpful and stimulating discussions with X. Crispin, C. Cramer and M. Sch\"onhoff are gratefully acknowledged. This project was supported by the French National Research Agency through grant ANR-19-CE30-0012-01 and by the European Research Council (ERC) through grant n$^\mathrm{o}$ 772725.


\bibliography{../../-Archive/literature}

\begin{thebibliography}{44}%
\makeatletter
\providecommand \@ifxundefined [1]{%
 \@ifx{#1\undefined}
}%
\providecommand \@ifnum [1]{%
 \ifnum #1\expandafter \@firstoftwo
 \else \expandafter \@secondoftwo
 \fi
}%
\providecommand \@ifx [1]{%
 \ifx #1\expandafter \@firstoftwo
 \else \expandafter \@secondoftwo
 \fi
}%
\providecommand \natexlab [1]{#1}%
\providecommand \enquote  [1]{``#1''}%
\providecommand \bibnamefont  [1]{#1}%
\providecommand \bibfnamefont [1]{#1}%
\providecommand \citenamefont [1]{#1}%
\providecommand \href@noop [0]{\@secondoftwo}%
\providecommand \href [0]{\begingroup \@sanitize@url \@href}%
\providecommand \@href[1]{\@@startlink{#1}\@@href}%
\providecommand \@@href[1]{\endgroup#1\@@endlink}%
\providecommand \@sanitize@url [0]{\catcode `\\12\catcode `\$12\catcode
  `\&12\catcode `\#12\catcode `\^12\catcode `\_12\catcode `\%12\relax}%
\providecommand \@@startlink[1]{}%
\providecommand \@@endlink[0]{}%
\providecommand \url  [0]{\begingroup\@sanitize@url \@url }%
\providecommand \@url [1]{\endgroup\@href {#1}{\urlprefix }}%
\providecommand \urlprefix  [0]{URL }%
\providecommand \Eprint [0]{\href }%
\providecommand \doibase [0]{http://dx.doi.org/}%
\providecommand \selectlanguage [0]{\@gobble}%
\providecommand \bibinfo  [0]{\@secondoftwo}%
\providecommand \bibfield  [0]{\@secondoftwo}%
\providecommand \translation [1]{[#1]}%
\providecommand \BibitemOpen [0]{}%
\providecommand \bibitemStop [0]{}%
\providecommand \bibitemNoStop [0]{.\EOS\space}%
\providecommand \EOS [0]{\spacefactor3000\relax}%
\providecommand \BibitemShut  [1]{\csname bibitem#1\endcsname}%
\let\auto@bib@innerbib\@empty
\bibitem [{\citenamefont {Bell}(2008)}]{Bell2008}%
  \BibitemOpen
  \bibfield  {author} {\bibinfo {author} {\bibfnamefont {L.~E.}\ \bibnamefont
  {Bell}},\ }\href {\doibase 10.1126/science.1158899} {\bibfield  {journal}
  {\bibinfo  {journal} {Science}\ }\textbf {\bibinfo {volume} {321}},\ \bibinfo
  {pages} {1457} (\bibinfo {year} {2008})}\BibitemShut {NoStop}%
\bibitem [{\citenamefont {Bubnova}\ \emph {et~al.}(2011)\citenamefont
  {Bubnova}, \citenamefont {Khan}, \citenamefont {Malti}, \citenamefont
  {Braun}, \citenamefont {Fahlman}, \citenamefont {Berggren},\ and\
  \citenamefont {Crispin}}]{Bubnova2011}%
  \BibitemOpen
  \bibfield  {author} {\bibinfo {author} {\bibfnamefont {O.}~\bibnamefont
  {Bubnova}}, \bibinfo {author} {\bibfnamefont {Z.~U.}\ \bibnamefont {Khan}},
  \bibinfo {author} {\bibfnamefont {A.}~\bibnamefont {Malti}}, \bibinfo
  {author} {\bibfnamefont {S.}~\bibnamefont {Braun}}, \bibinfo {author}
  {\bibfnamefont {M.}~\bibnamefont {Fahlman}}, \bibinfo {author} {\bibfnamefont
  {M.}~\bibnamefont {Berggren}}, \ and\ \bibinfo {author} {\bibfnamefont
  {X.}~\bibnamefont {Crispin}},\ }\href {\doibase 10.1038/nmat3012} {\bibfield
  {journal} {\bibinfo  {journal} {Nature Materials}\ }\textbf {\bibinfo
  {volume} {10}},\ \bibinfo {pages} {429} (\bibinfo {year} {2011})}\BibitemShut
  {NoStop}%
\bibitem [{\citenamefont {Russ}\ \emph {et~al.}(2016)\citenamefont {Russ},
  \citenamefont {Glaudell}, \citenamefont {Urban}, \citenamefont {Chabinyc},\
  and\ \citenamefont {Segalman}}]{Russ2016}%
  \BibitemOpen
  \bibfield  {author} {\bibinfo {author} {\bibfnamefont {B.}~\bibnamefont
  {Russ}}, \bibinfo {author} {\bibfnamefont {A.}~\bibnamefont {Glaudell}},
  \bibinfo {author} {\bibfnamefont {J.~J.}\ \bibnamefont {Urban}}, \bibinfo
  {author} {\bibfnamefont {M.~L.}\ \bibnamefont {Chabinyc}}, \ and\ \bibinfo
  {author} {\bibfnamefont {R.~A.}\ \bibnamefont {Segalman}},\ }\href {\doibase
  10.1038/natrevmats.2016.50} {\bibfield  {journal} {\bibinfo  {journal}
  {Nature Reviews Materials}\ }\textbf {\bibinfo {volume} {1}},\ \bibinfo
  {pages} {16050} (\bibinfo {year} {2016})}\BibitemShut {NoStop}%
\bibitem [{\citenamefont {Gregory}\ \emph {et~al.}(2018)\citenamefont
  {Gregory}, \citenamefont {Menon}, \citenamefont {Ye}, \citenamefont
  {Seferos}, \citenamefont {Reynolds},\ and\ \citenamefont
  {Yee}}]{Gregory2018}%
  \BibitemOpen
  \bibfield  {author} {\bibinfo {author} {\bibfnamefont {S.~A.}\ \bibnamefont
  {Gregory}}, \bibinfo {author} {\bibfnamefont {A.~K.}\ \bibnamefont {Menon}},
  \bibinfo {author} {\bibfnamefont {S.}~\bibnamefont {Ye}}, \bibinfo {author}
  {\bibfnamefont {D.~S.}\ \bibnamefont {Seferos}}, \bibinfo {author}
  {\bibfnamefont {J.~R.}\ \bibnamefont {Reynolds}}, \ and\ \bibinfo {author}
  {\bibfnamefont {S.~K.}\ \bibnamefont {Yee}},\ }\href {\doibase
  10.1002/aenm.201802419} {\bibfield  {journal} {\bibinfo  {journal} {Advanced
  Energy Materials}\ }\textbf {\bibinfo {volume} {8}},\ \bibinfo {pages}
  {1802419} (\bibinfo {year} {2018})}\BibitemShut {NoStop}%
\bibitem [{\citenamefont {Xiao}\ \emph {et~al.}(2017)\citenamefont {Xiao},
  \citenamefont {Widenmeyer}, \citenamefont {Xie}, \citenamefont {Zou},
  \citenamefont {Yoon}, \citenamefont {Scavini}, \citenamefont {Checchia},
  \citenamefont {Zhong}, \citenamefont {Hansmann}, \citenamefont {Kilper},
  \citenamefont {Kovalevsky},\ and\ \citenamefont {Weidenkaff}}]{Xiao2017}%
  \BibitemOpen
  \bibfield  {author} {\bibinfo {author} {\bibfnamefont {X.}~\bibnamefont
  {Xiao}}, \bibinfo {author} {\bibfnamefont {M.}~\bibnamefont {Widenmeyer}},
  \bibinfo {author} {\bibfnamefont {W.}~\bibnamefont {Xie}}, \bibinfo {author}
  {\bibfnamefont {T.}~\bibnamefont {Zou}}, \bibinfo {author} {\bibfnamefont
  {S.}~\bibnamefont {Yoon}}, \bibinfo {author} {\bibfnamefont {M.}~\bibnamefont
  {Scavini}}, \bibinfo {author} {\bibfnamefont {S.}~\bibnamefont {Checchia}},
  \bibinfo {author} {\bibfnamefont {Z.}~\bibnamefont {Zhong}}, \bibinfo
  {author} {\bibfnamefont {P.}~\bibnamefont {Hansmann}}, \bibinfo {author}
  {\bibfnamefont {S.}~\bibnamefont {Kilper}}, \bibinfo {author} {\bibfnamefont
  {A.}~\bibnamefont {Kovalevsky}}, \ and\ \bibinfo {author} {\bibfnamefont
  {A.}~\bibnamefont {Weidenkaff}},\ }\href {\doibase 10.1039/c7cp00020k}
  {\bibfield  {journal} {\bibinfo  {journal} {Phys Chem Chem Phys}\ }\textbf
  {\bibinfo {volume} {19}},\ \bibinfo {pages} {13469} (\bibinfo {year}
  {2017})}\BibitemShut {NoStop}%
\bibitem [{\citenamefont {Zhao}\ \emph {et~al.}(2016)\citenamefont {Zhao},
  \citenamefont {Wang}, \citenamefont {Khan}, \citenamefont {Chen},
  \citenamefont {Gabrielsson}, \citenamefont {Jonsson}, \citenamefont
  {Berggren},\ and\ \citenamefont {Crispin}}]{Zhao2016}%
  \BibitemOpen
  \bibfield  {author} {\bibinfo {author} {\bibfnamefont {D.}~\bibnamefont
  {Zhao}}, \bibinfo {author} {\bibfnamefont {H.}~\bibnamefont {Wang}}, \bibinfo
  {author} {\bibfnamefont {Z.~U.}\ \bibnamefont {Khan}}, \bibinfo {author}
  {\bibfnamefont {J.~C.}\ \bibnamefont {Chen}}, \bibinfo {author}
  {\bibfnamefont {R.}~\bibnamefont {Gabrielsson}}, \bibinfo {author}
  {\bibfnamefont {M.~P.}\ \bibnamefont {Jonsson}}, \bibinfo {author}
  {\bibfnamefont {M.}~\bibnamefont {Berggren}}, \ and\ \bibinfo {author}
  {\bibfnamefont {X.}~\bibnamefont {Crispin}},\ }\href {\doibase
  10.1039/C6EE00121A} {\bibfield  {journal} {\bibinfo  {journal} {Energy
  Environ. Sci.}\ }\textbf {\bibinfo {volume} {9}},\ \bibinfo {pages} {1450}
  (\bibinfo {year} {2016})}\BibitemShut {NoStop}%
\bibitem [{\citenamefont {Kim}\ \emph {et~al.}(2016)\citenamefont {Kim},
  \citenamefont {Lin},\ and\ \citenamefont {Yu}}]{Kim2016}%
  \BibitemOpen
  \bibfield  {author} {\bibinfo {author} {\bibfnamefont {S.~L.}\ \bibnamefont
  {Kim}}, \bibinfo {author} {\bibfnamefont {H.~T.}\ \bibnamefont {Lin}}, \ and\
  \bibinfo {author} {\bibfnamefont {C.}~\bibnamefont {Yu}},\ }\href {\doibase
  10.1002/aenm.201600546} {\bibfield  {journal} {\bibinfo  {journal} {Advanced
  Energy Materials}\ }\textbf {\bibinfo {volume} {6}},\ \bibinfo {pages}
  {1600546} (\bibinfo {year} {2016})}\BibitemShut {NoStop}%
\bibitem [{\citenamefont {Wang}\ \emph {et~al.}(2017)\citenamefont {Wang},
  \citenamefont {Zhao}, \citenamefont {Khan}, \citenamefont {Puzinas},
  \citenamefont {Jonsson}, \citenamefont {Berggren},\ and\ \citenamefont
  {Crispin}}]{Wang2017}%
  \BibitemOpen
  \bibfield  {author} {\bibinfo {author} {\bibfnamefont {H.}~\bibnamefont
  {Wang}}, \bibinfo {author} {\bibfnamefont {D.}~\bibnamefont {Zhao}}, \bibinfo
  {author} {\bibfnamefont {Z.~U.}\ \bibnamefont {Khan}}, \bibinfo {author}
  {\bibfnamefont {S.}~\bibnamefont {Puzinas}}, \bibinfo {author} {\bibfnamefont
  {M.~P.}\ \bibnamefont {Jonsson}}, \bibinfo {author} {\bibfnamefont
  {M.}~\bibnamefont {Berggren}}, \ and\ \bibinfo {author} {\bibfnamefont
  {X.}~\bibnamefont {Crispin}},\ }\href {\doibase 10.1002/aelm.201700013}
  {\bibfield  {journal} {\bibinfo  {journal} {Advanced Electronic Materials}\
  }\textbf {\bibinfo {volume} {3}},\ \bibinfo {pages} {1700013} (\bibinfo
  {year} {2017})}\BibitemShut {NoStop}%
\bibitem [{\citenamefont {Li}\ \emph {et~al.}(2019)\citenamefont {Li},
  \citenamefont {Zhang}, \citenamefont {Lacey}, \citenamefont {Mi},
  \citenamefont {Zhao}, \citenamefont {Jiang}, \citenamefont {Song},
  \citenamefont {Liu}, \citenamefont {Chen}, \citenamefont {Dai}, \citenamefont
  {Yao}, \citenamefont {Das}, \citenamefont {Yang}, \citenamefont {Briber},\
  and\ \citenamefont {Hu}}]{Li2019}%
  \BibitemOpen
  \bibfield  {author} {\bibinfo {author} {\bibfnamefont {T.}~\bibnamefont
  {Li}}, \bibinfo {author} {\bibfnamefont {X.}~\bibnamefont {Zhang}}, \bibinfo
  {author} {\bibfnamefont {S.~D.}\ \bibnamefont {Lacey}}, \bibinfo {author}
  {\bibfnamefont {R.}~\bibnamefont {Mi}}, \bibinfo {author} {\bibfnamefont
  {X.}~\bibnamefont {Zhao}}, \bibinfo {author} {\bibfnamefont {F.}~\bibnamefont
  {Jiang}}, \bibinfo {author} {\bibfnamefont {J.}~\bibnamefont {Song}},
  \bibinfo {author} {\bibfnamefont {Z.}~\bibnamefont {Liu}}, \bibinfo {author}
  {\bibfnamefont {G.}~\bibnamefont {Chen}}, \bibinfo {author} {\bibfnamefont
  {J.}~\bibnamefont {Dai}}, \bibinfo {author} {\bibfnamefont {Y.}~\bibnamefont
  {Yao}}, \bibinfo {author} {\bibfnamefont {S.}~\bibnamefont {Das}}, \bibinfo
  {author} {\bibfnamefont {R.}~\bibnamefont {Yang}}, \bibinfo {author}
  {\bibfnamefont {R.~M.}\ \bibnamefont {Briber}}, \ and\ \bibinfo {author}
  {\bibfnamefont {L.}~\bibnamefont {Hu}},\ }\href {\doibase
  10.1038/s41563-019-0315-6} {\bibfield  {journal} {\bibinfo  {journal} {Nature
  Materials}\ }\textbf {\bibinfo {volume} {18}},\ \bibinfo {pages} {608}
  (\bibinfo {year} {2019})}\BibitemShut {NoStop}%
\bibitem [{\citenamefont {Zhao}\ \emph {et~al.}(2019)\citenamefont {Zhao},
  \citenamefont {Martinelli}, \citenamefont {Willfahrt}, \citenamefont
  {Fischer}, \citenamefont {Bernin}, \citenamefont {Khan}, \citenamefont
  {Shahi}, \citenamefont {Brill}, \citenamefont {Jonsson}, \citenamefont
  {Fabiano},\ and\ \citenamefont {Crispin}}]{Zhao2019}%
  \BibitemOpen
  \bibfield  {author} {\bibinfo {author} {\bibfnamefont {D.}~\bibnamefont
  {Zhao}}, \bibinfo {author} {\bibfnamefont {A.}~\bibnamefont {Martinelli}},
  \bibinfo {author} {\bibfnamefont {A.}~\bibnamefont {Willfahrt}}, \bibinfo
  {author} {\bibfnamefont {T.}~\bibnamefont {Fischer}}, \bibinfo {author}
  {\bibfnamefont {D.}~\bibnamefont {Bernin}}, \bibinfo {author} {\bibfnamefont
  {Z.~U.}\ \bibnamefont {Khan}}, \bibinfo {author} {\bibfnamefont
  {M.}~\bibnamefont {Shahi}}, \bibinfo {author} {\bibfnamefont
  {J.}~\bibnamefont {Brill}}, \bibinfo {author} {\bibfnamefont {M.~P.}\
  \bibnamefont {Jonsson}}, \bibinfo {author} {\bibfnamefont {S.}~\bibnamefont
  {Fabiano}}, \ and\ \bibinfo {author} {\bibfnamefont {X.}~\bibnamefont
  {Crispin}},\ }\href {\doibase 10.1038/s41467-019-08930-7} {\bibfield
  {journal} {\bibinfo  {journal} {Nature Communications}\ }\textbf {\bibinfo
  {volume} {10}},\ \bibinfo {pages} {1093} (\bibinfo {year}
  {2019})}\BibitemShut {NoStop}%
\bibitem [{\citenamefont {Han}\ \emph {et~al.}(2020{\natexlab{a}})\citenamefont
  {Han}, \citenamefont {Qian}, \citenamefont {Li}, \citenamefont {Deng},
  \citenamefont {Zhu}, \citenamefont {Han}, \citenamefont {Zhang},
  \citenamefont {Wang}, \citenamefont {Feng}, \citenamefont {Chen},\ and\
  \citenamefont {Liu}}]{Han2020}%
  \BibitemOpen
  \bibfield  {author} {\bibinfo {author} {\bibfnamefont {C.-G.}\ \bibnamefont
  {Han}}, \bibinfo {author} {\bibfnamefont {X.}~\bibnamefont {Qian}}, \bibinfo
  {author} {\bibfnamefont {Q.}~\bibnamefont {Li}}, \bibinfo {author}
  {\bibfnamefont {B.}~\bibnamefont {Deng}}, \bibinfo {author} {\bibfnamefont
  {Y.}~\bibnamefont {Zhu}}, \bibinfo {author} {\bibfnamefont {Z.}~\bibnamefont
  {Han}}, \bibinfo {author} {\bibfnamefont {W.}~\bibnamefont {Zhang}}, \bibinfo
  {author} {\bibfnamefont {W.}~\bibnamefont {Wang}}, \bibinfo {author}
  {\bibfnamefont {S.-P.}\ \bibnamefont {Feng}}, \bibinfo {author}
  {\bibfnamefont {G.}~\bibnamefont {Chen}}, \ and\ \bibinfo {author}
  {\bibfnamefont {W.}~\bibnamefont {Liu}},\ }\href {\doibase
  10.1126/science.aaz5045} {\bibfield  {journal} {\bibinfo  {journal}
  {Science}\ }\textbf {\bibinfo {volume} {368}},\ \bibinfo {pages} {1091}
  (\bibinfo {year} {2020}{\natexlab{a}})}\BibitemShut {NoStop}%
\bibitem [{\citenamefont {Wang}\ \emph {et~al.}(2015)\citenamefont {Wang},
  \citenamefont {Ail}, \citenamefont {Gabrielsson}, \citenamefont {Berggren},\
  and\ \citenamefont {Crispin}}]{Wang2015}%
  \BibitemOpen
  \bibfield  {author} {\bibinfo {author} {\bibfnamefont {H.}~\bibnamefont
  {Wang}}, \bibinfo {author} {\bibfnamefont {U.}~\bibnamefont {Ail}}, \bibinfo
  {author} {\bibfnamefont {R.}~\bibnamefont {Gabrielsson}}, \bibinfo {author}
  {\bibfnamefont {M.}~\bibnamefont {Berggren}}, \ and\ \bibinfo {author}
  {\bibfnamefont {X.}~\bibnamefont {Crispin}},\ }\href {\doibase
  10.1002/aenm.201500044} {\bibfield  {journal} {\bibinfo  {journal} {Advanced
  Energy Materials}\ }\textbf {\bibinfo {volume} {5}},\ \bibinfo {pages}
  {1500044} (\bibinfo {year} {2015})}\BibitemShut {NoStop}%
\bibitem [{\citenamefont {Chang}\ \emph {et~al.}(2016)\citenamefont {Chang},
  \citenamefont {Fang}, \citenamefont {Liu}, \citenamefont {Evans},
  \citenamefont {Russ}, \citenamefont {Popere}, \citenamefont {Patel},
  \citenamefont {Chabinyc},\ and\ \citenamefont {Segalman}}]{Chang2016}%
  \BibitemOpen
  \bibfield  {author} {\bibinfo {author} {\bibfnamefont {W.~B.}\ \bibnamefont
  {Chang}}, \bibinfo {author} {\bibfnamefont {H.}~\bibnamefont {Fang}},
  \bibinfo {author} {\bibfnamefont {J.}~\bibnamefont {Liu}}, \bibinfo {author}
  {\bibfnamefont {C.~M.}\ \bibnamefont {Evans}}, \bibinfo {author}
  {\bibfnamefont {B.}~\bibnamefont {Russ}}, \bibinfo {author} {\bibfnamefont
  {B.~C.}\ \bibnamefont {Popere}}, \bibinfo {author} {\bibfnamefont {S.~N.}\
  \bibnamefont {Patel}}, \bibinfo {author} {\bibfnamefont {M.~L.}\ \bibnamefont
  {Chabinyc}}, \ and\ \bibinfo {author} {\bibfnamefont {R.~A.}\ \bibnamefont
  {Segalman}},\ }\bibfield  {booktitle} {\emph {\bibinfo {booktitle} {ACS Macro
  Letters}},\ }\href {\doibase 10.1021/acsmacrolett.6b00054} {\bibfield
  {journal} {\bibinfo  {journal} {ACS Macro Letters}\ }\textbf {\bibinfo
  {volume} {5}},\ \bibinfo {pages} {455} (\bibinfo {year} {2016})}\BibitemShut
  {NoStop}%
\bibitem [{\citenamefont {de~Groot}\ and\ \citenamefont
  {Mazur}(1962)}]{deGroot1962}%
  \BibitemOpen
  \bibfield  {author} {\bibinfo {author} {\bibfnamefont {S.}~\bibnamefont
  {de~Groot}}\ and\ \bibinfo {author} {\bibfnamefont {P.}~\bibnamefont
  {Mazur}},\ }\href@noop {} {\emph {\bibinfo {title} {Non-equilibrium
  Thermodynamics}}}\ (\bibinfo  {publisher} {North-Holland Publishing Company;
  Interscience Publishers},\ \bibinfo {year} {1962})\BibitemShut {NoStop}%
\bibitem [{\citenamefont {Fritzsche}(1971)}]{Fritsche1971}%
  \BibitemOpen
  \bibfield  {author} {\bibinfo {author} {\bibfnamefont {H.}~\bibnamefont
  {Fritzsche}},\ }\href {\doibase https://doi.org/10.1016/0038-1098(71)90096-2}
  {\bibfield  {journal} {\bibinfo  {journal} {Solid State Communications}\
  }\textbf {\bibinfo {volume} {9}},\ \bibinfo {pages} {1813 } (\bibinfo {year}
  {1971})}\BibitemShut {NoStop}%
\bibitem [{\citenamefont {Mott}\ and\ \citenamefont {Davis}(1971)}]{Mott1971}%
  \BibitemOpen
  \bibfield  {author} {\bibinfo {author} {\bibfnamefont {N.~F.}\ \bibnamefont
  {Mott}}\ and\ \bibinfo {author} {\bibfnamefont {E.~A.}\ \bibnamefont
  {Davis}},\ }\href@noop {} {\emph {\bibinfo {title} {Electronic Processes in
  Non-Crystalline Materials}}}\ (\bibinfo  {publisher} {Oxford University
  Press},\ \bibinfo {year} {1971})\BibitemShut {NoStop}%
\bibitem [{\citenamefont {Nernst}(1889)}]{Nernst1889}%
  \BibitemOpen
  \bibfield  {author} {\bibinfo {author} {\bibfnamefont {W.}~\bibnamefont
  {Nernst}},\ }\href@noop {} {\bibfield  {journal} {\bibinfo  {journal}
  {Zeitschrift f{\"u}r physikalische Chemie}\ }\textbf {\bibinfo {volume}
  {4}},\ \bibinfo {pages} {129} (\bibinfo {year} {1889})}\BibitemShut {NoStop}%
\bibitem [{\citenamefont {Agar}\ \emph {et~al.}(1989)\citenamefont {Agar},
  \citenamefont {Mou},\ and\ \citenamefont {Lin}}]{Agar1989}%
  \BibitemOpen
  \bibfield  {author} {\bibinfo {author} {\bibfnamefont {J.~N.}\ \bibnamefont
  {Agar}}, \bibinfo {author} {\bibfnamefont {C.~Y.}\ \bibnamefont {Mou}}, \
  and\ \bibinfo {author} {\bibfnamefont {J.~L.}\ \bibnamefont {Lin}},\ }\href
  {\doibase 10.1021/j100342a073} {\bibfield  {journal} {\bibinfo  {journal}
  {The Journal of Physical Chemistry}\ }\textbf {\bibinfo {volume} {93}},\
  \bibinfo {pages} {2079} (\bibinfo {year} {1989})}\BibitemShut {NoStop}%
\bibitem [{\citenamefont {Putnam}\ and\ \citenamefont
  {Cahill}(2005)}]{Putnam2005}%
  \BibitemOpen
  \bibfield  {author} {\bibinfo {author} {\bibfnamefont {S.~A.}\ \bibnamefont
  {Putnam}}\ and\ \bibinfo {author} {\bibfnamefont {D.~G.}\ \bibnamefont
  {Cahill}},\ }\href@noop {} {\bibfield  {journal} {\bibinfo  {journal}
  {Langmuir}\ }\textbf {\bibinfo {volume} {21}},\ \bibinfo {pages} {5317}
  (\bibinfo {year} {2005})}\BibitemShut {NoStop}%
\bibitem [{\citenamefont {W\"urger}(2010)}]{Wuerger2010}%
  \BibitemOpen
  \bibfield  {author} {\bibinfo {author} {\bibfnamefont {A.}~\bibnamefont
  {W\"urger}},\ }\href {\doibase 10.1088/0034-4885/73/12/126601} {\bibfield
  {journal} {\bibinfo  {journal} {Rep. Prog. Phys.}\ }\textbf {\bibinfo
  {volume} {73}},\ \bibinfo {pages} {126601} (\bibinfo {year}
  {2010})}\BibitemShut {NoStop}%
\bibitem [{\citenamefont {Vigolo}\ \emph {et~al.}(2010)\citenamefont {Vigolo},
  \citenamefont {Buzzaccaro},\ and\ \citenamefont {Piazza}}]{Vigolo2010}%
  \BibitemOpen
  \bibfield  {author} {\bibinfo {author} {\bibfnamefont {D.}~\bibnamefont
  {Vigolo}}, \bibinfo {author} {\bibfnamefont {S.}~\bibnamefont {Buzzaccaro}},
  \ and\ \bibinfo {author} {\bibfnamefont {R.}~\bibnamefont {Piazza}},\ }\href
  {\doibase 10.1021/la904588s} {\bibfield  {journal} {\bibinfo  {journal}
  {Langmuir}\ }\textbf {\bibinfo {volume} {26}},\ \bibinfo {pages} {7792}
  (\bibinfo {year} {2010})}\BibitemShut {NoStop}%
\bibitem [{\citenamefont {Majee}\ and\ \citenamefont
  {W\"urger}(2011)}]{Majee2011}%
  \BibitemOpen
  \bibfield  {author} {\bibinfo {author} {\bibfnamefont {A.}~\bibnamefont
  {Majee}}\ and\ \bibinfo {author} {\bibfnamefont {A.}~\bibnamefont
  {W\"urger}},\ }\href {\doibase 10.1103/PhysRevE.83.061403} {\bibfield
  {journal} {\bibinfo  {journal} {Phys. Rev. E}\ }\textbf {\bibinfo {volume}
  {83}},\ \bibinfo {pages} {061403} (\bibinfo {year} {2011})}\BibitemShut
  {NoStop}%
\bibitem [{\citenamefont {Eslahian}\ \emph {et~al.}(2014)\citenamefont
  {Eslahian}, \citenamefont {Majee}, \citenamefont {Maskos},\ and\
  \citenamefont {W{\"u}rger}}]{Eslahian2014}%
  \BibitemOpen
  \bibfield  {author} {\bibinfo {author} {\bibfnamefont {K.~A.}\ \bibnamefont
  {Eslahian}}, \bibinfo {author} {\bibfnamefont {A.}~\bibnamefont {Majee}},
  \bibinfo {author} {\bibfnamefont {M.}~\bibnamefont {Maskos}}, \ and\ \bibinfo
  {author} {\bibfnamefont {A.}~\bibnamefont {W{\"u}rger}},\ }\href@noop {}
  {\bibfield  {journal} {\bibinfo  {journal} {Soft Matter}\ }\textbf {\bibinfo
  {volume} {10}},\ \bibinfo {pages} {1931} (\bibinfo {year}
  {2014})}\BibitemShut {NoStop}%
\bibitem [{\citenamefont {Eyring}(1935)}]{Eyring1935}%
  \BibitemOpen
  \bibfield  {author} {\bibinfo {author} {\bibfnamefont {H.}~\bibnamefont
  {Eyring}},\ }\href@noop {} {\bibfield  {journal} {\bibinfo  {journal}
  {Chemical Reviews}\ }\textbf {\bibinfo {volume} {17}},\ \bibinfo {pages} {65}
  (\bibinfo {year} {1935})}\BibitemShut {NoStop}%
\bibitem [{\citenamefont {De}\ \emph {et~al.}(2017)\citenamefont {De},
  \citenamefont {Ostendorf}, \citenamefont {Sch{\"o}nhoff},\ and\ \citenamefont
  {Cramer}}]{De2017}%
  \BibitemOpen
  \bibfield  {author} {\bibinfo {author} {\bibfnamefont {S.}~\bibnamefont
  {De}}, \bibinfo {author} {\bibfnamefont {A.}~\bibnamefont {Ostendorf}},
  \bibinfo {author} {\bibfnamefont {M.}~\bibnamefont {Sch{\"o}nhoff}}, \ and\
  \bibinfo {author} {\bibfnamefont {C.}~\bibnamefont {Cramer}},\ }\href@noop {}
  {\bibfield  {journal} {\bibinfo  {journal} {Polymers}\ }\textbf {\bibinfo
  {volume} {9}},\ \bibinfo {pages} {550} (\bibinfo {year} {2017})}\BibitemShut
  {NoStop}%
\bibitem [{\citenamefont {Ostendorf}\ \emph {et~al.}(2019)\citenamefont
  {Ostendorf}, \citenamefont {Sch{\"o}nhoff},\ and\ \citenamefont
  {Cramer}}]{Ostendorf2019}%
  \BibitemOpen
  \bibfield  {author} {\bibinfo {author} {\bibfnamefont {A.}~\bibnamefont
  {Ostendorf}}, \bibinfo {author} {\bibfnamefont {M.}~\bibnamefont
  {Sch{\"o}nhoff}}, \ and\ \bibinfo {author} {\bibfnamefont {C.}~\bibnamefont
  {Cramer}},\ }\href@noop {} {\bibfield  {journal} {\bibinfo  {journal}
  {Physical Chemistry Chemical Physics}\ }\textbf {\bibinfo {volume} {21}},\
  \bibinfo {pages} {7321} (\bibinfo {year} {2019})}\BibitemShut {NoStop}%
\bibitem [{SI-()}]{SI-TE2020}%
  \BibitemOpen
  \href@noop {} {\bibinfo  {journal} {The supplementary information file
  discusses the case of finite bias energy and the linearization
  approximation}\ }\BibitemShut {NoStop}%
\bibitem [{\citenamefont {LeClaire}(1954)}]{LeClaire1954}%
  \BibitemOpen
\bibfield  {journal} {  }\bibfield  {author} {\bibinfo {author} {\bibfnamefont
  {A.~D.}\ \bibnamefont {LeClaire}},\ }\href {\doibase 10.1103/PhysRev.93.344}
  {\bibfield  {journal} {\bibinfo  {journal} {Phys. Rev.}\ }\textbf {\bibinfo
  {volume} {93}},\ \bibinfo {pages} {344} (\bibinfo {year} {1954})}\BibitemShut
  {NoStop}%
\bibitem [{\citenamefont {Brinkman}(1954)}]{Brinkman1954}%
  \BibitemOpen
  \bibfield  {author} {\bibinfo {author} {\bibfnamefont {J.~A.}\ \bibnamefont
  {Brinkman}},\ }\href {\doibase 10.1103/PhysRev.93.345} {\bibfield  {journal}
  {\bibinfo  {journal} {Phys. Rev.}\ }\textbf {\bibinfo {volume} {93}},\
  \bibinfo {pages} {345} (\bibinfo {year} {1954})}\BibitemShut {NoStop}%
\bibitem [{\citenamefont {Abbott}(2004)}]{Abbott2004}%
  \BibitemOpen
  \bibfield  {author} {\bibinfo {author} {\bibfnamefont {A.~P.}\ \bibnamefont
  {Abbott}},\ }\href {\doibase 10.1002/cphc.200400190} {\bibfield  {journal}
  {\bibinfo  {journal} {ChemPhysChem}\ }\textbf {\bibinfo {volume} {5}},\
  \bibinfo {pages} {1242} (\bibinfo {year} {2004})}\BibitemShut {NoStop}%
\bibitem [{\citenamefont {Jaffe}\ and\ \citenamefont
  {Shewmon}(1964)}]{Jaffe1964}%
  \BibitemOpen
  \bibfield  {author} {\bibinfo {author} {\bibfnamefont {D.}~\bibnamefont
  {Jaffe}}\ and\ \bibinfo {author} {\bibfnamefont {P.}~\bibnamefont
  {Shewmon}},\ }\href {\doibase https://doi.org/10.1016/0001-6160(64)90024-0}
  {\bibfield  {journal} {\bibinfo  {journal} {Acta Metallurgica}\ }\textbf
  {\bibinfo {volume} {12}},\ \bibinfo {pages} {515 } (\bibinfo {year}
  {1964})}\BibitemShut {NoStop}%
\bibitem [{\citenamefont {Han}\ \emph {et~al.}(2020{\natexlab{b}})\citenamefont
  {Han}, \citenamefont {Cheon}, \citenamefont {Kim}, \citenamefont {Lee},
  \citenamefont {Kim},\ and\ \citenamefont {Jung}}]{Han2020a}%
  \BibitemOpen
  \bibfield  {author} {\bibinfo {author} {\bibfnamefont {Y.~K.}\ \bibnamefont
  {Han}}, \bibinfo {author} {\bibfnamefont {J.~Y.}\ \bibnamefont {Cheon}},
  \bibinfo {author} {\bibfnamefont {T.}~\bibnamefont {Kim}}, \bibinfo {author}
  {\bibfnamefont {S.~B.}\ \bibnamefont {Lee}}, \bibinfo {author} {\bibfnamefont
  {Y.~D.}\ \bibnamefont {Kim}}, \ and\ \bibinfo {author} {\bibfnamefont
  {B.~M.}\ \bibnamefont {Jung}},\ }\href {\doibase 10.1039/D0RA02327B}
  {\bibfield  {journal} {\bibinfo  {journal} {RSC Adv.}\ }\textbf {\bibinfo
  {volume} {10}},\ \bibinfo {pages} {18945} (\bibinfo {year}
  {2020}{\natexlab{b}})}\BibitemShut {NoStop}%
\bibitem [{\citenamefont {D'Agostino}\ \emph {et~al.}(2018)\citenamefont
  {D'Agostino}, \citenamefont {Mantle}, \citenamefont {Mullan}, \citenamefont
  {Hardacre},\ and\ \citenamefont {Gladden}}]{Agostino2018}%
  \BibitemOpen
  \bibfield  {author} {\bibinfo {author} {\bibfnamefont {C.}~\bibnamefont
  {D'Agostino}}, \bibinfo {author} {\bibfnamefont {M.~D.}\ \bibnamefont
  {Mantle}}, \bibinfo {author} {\bibfnamefont {C.~L.}\ \bibnamefont {Mullan}},
  \bibinfo {author} {\bibfnamefont {C.}~\bibnamefont {Hardacre}}, \ and\
  \bibinfo {author} {\bibfnamefont {L.~F.}\ \bibnamefont {Gladden}},\ }\href
  {\doibase 10.1002/cphc.201701354} {\bibfield  {journal} {\bibinfo  {journal}
  {ChemPhysChem}\ }\textbf {\bibinfo {volume} {19}},\ \bibinfo {pages} {1081}
  (\bibinfo {year} {2018})}\BibitemShut {NoStop}%
\bibitem [{\citenamefont {Tafur}\ \emph {et~al.}(2015)\citenamefont {Tafur},
  \citenamefont {Santos},\ and\ \citenamefont {Fernandez~Romero}}]{Tafur2015}%
  \BibitemOpen
  \bibfield  {author} {\bibinfo {author} {\bibfnamefont {J.~P.}\ \bibnamefont
  {Tafur}}, \bibinfo {author} {\bibfnamefont {F.}~\bibnamefont {Santos}}, \
  and\ \bibinfo {author} {\bibfnamefont {A.~J.}\ \bibnamefont
  {Fernandez~Romero}},\ }\href@noop {} {\bibfield  {journal} {\bibinfo
  {journal} {Membranes}\ }\textbf {\bibinfo {volume} {5}},\ \bibinfo {pages}
  {752} (\bibinfo {year} {2015})}\BibitemShut {NoStop}%
\bibitem [{\citenamefont {Michaels}\ \emph {et~al.}(1965)\citenamefont
  {Michaels}, \citenamefont {Falkenstein},\ and\ \citenamefont
  {Schneider}}]{Michaels1965}%
  \BibitemOpen
  \bibfield  {author} {\bibinfo {author} {\bibfnamefont {A.~S.}\ \bibnamefont
  {Michaels}}, \bibinfo {author} {\bibfnamefont {G.~L.}\ \bibnamefont
  {Falkenstein}}, \ and\ \bibinfo {author} {\bibfnamefont {N.~S.}\ \bibnamefont
  {Schneider}},\ }\href@noop {} {\bibfield  {journal} {\bibinfo  {journal} {The
  Journal of Physical Chemistry}\ }\textbf {\bibinfo {volume} {69}},\ \bibinfo
  {pages} {1456} (\bibinfo {year} {1965})}\BibitemShut {NoStop}%
\bibitem [{\citenamefont {Varcoe}(2007)}]{Varcoe2007}%
  \BibitemOpen
  \bibfield  {author} {\bibinfo {author} {\bibfnamefont {J.~R.}\ \bibnamefont
  {Varcoe}},\ }\href@noop {} {\bibfield  {journal} {\bibinfo  {journal}
  {Physical Chemistry Chemical Physics}\ }\textbf {\bibinfo {volume} {9}},\
  \bibinfo {pages} {1479} (\bibinfo {year} {2007})}\BibitemShut {NoStop}%
\bibitem [{\citenamefont {Wiegand}(2004)}]{Wiegand2004}%
  \BibitemOpen
  \bibfield  {author} {\bibinfo {author} {\bibfnamefont {S.}~\bibnamefont
  {Wiegand}},\ }\href@noop {} {\bibfield  {journal} {\bibinfo  {journal} {J.
  Phys.: Condens. Matter}\ }\textbf {\bibinfo {volume} {16}},\ \bibinfo {pages}
  {R357} (\bibinfo {year} {2004})}\BibitemShut {NoStop}%
\bibitem [{\citenamefont {Morthomas}\ and\ \citenamefont
  {W\"urger}(2008)}]{Morthomas2008}%
  \BibitemOpen
  \bibfield  {author} {\bibinfo {author} {\bibfnamefont {J.}~\bibnamefont
  {Morthomas}}\ and\ \bibinfo {author} {\bibfnamefont {A.}~\bibnamefont
  {W\"urger}},\ }\href@noop {} {\bibfield  {journal} {\bibinfo  {journal} {Eur.
  Phys. J. E}\ }\textbf {\bibinfo {volume} {27}} (\bibinfo {year}
  {2008})}\BibitemShut {NoStop}%
\bibitem [{\citenamefont {F{\"u}rth}(1941)}]{Fuerth1941}%
  \BibitemOpen
  \bibfield  {author} {\bibinfo {author} {\bibfnamefont {R.}~\bibnamefont
  {F{\"u}rth}},\ }\href {\doibase 10.1017/S0305004100021757} {\bibfield
  {journal} {\bibinfo  {journal} {Mathematical Proceedings of the Cambridge
  Philosophical Society}\ }\textbf {\bibinfo {volume} {37}},\ \bibinfo {pages}
  {276} (\bibinfo {year} {1941})}\BibitemShut {NoStop}%
\bibitem [{\citenamefont {Kramers}(1940)}]{Kramers1940}%
  \BibitemOpen
  \bibfield  {author} {\bibinfo {author} {\bibfnamefont {H.~A.}\ \bibnamefont
  {Kramers}},\ }\href@noop {} {\bibfield  {journal} {\bibinfo  {journal}
  {Physica}\ }\textbf {\bibinfo {volume} {7}},\ \bibinfo {pages} {284}
  (\bibinfo {year} {1940})}\BibitemShut {NoStop}%
\bibitem [{\citenamefont {Jiang}\ \emph {et~al.}(2009)\citenamefont {Jiang},
  \citenamefont {Wada}, \citenamefont {Yoshinaga},\ and\ \citenamefont
  {Sano}}]{Jiang2009}%
  \BibitemOpen
  \bibfield  {author} {\bibinfo {author} {\bibfnamefont {H.-R.}\ \bibnamefont
  {Jiang}}, \bibinfo {author} {\bibfnamefont {H.}~\bibnamefont {Wada}},
  \bibinfo {author} {\bibfnamefont {N.}~\bibnamefont {Yoshinaga}}, \ and\
  \bibinfo {author} {\bibfnamefont {M.}~\bibnamefont {Sano}},\ }\href {\doibase
  10.1103/PhysRevLett.102.208301} {\bibfield  {journal} {\bibinfo  {journal}
  {Phys. Rev. Lett.}\ }\textbf {\bibinfo {volume} {102}},\ \bibinfo {pages}
  {208301} (\bibinfo {year} {2009})}\BibitemShut {NoStop}%
\bibitem [{\citenamefont {Buttinoni}\ \emph {et~al.}(2012)\citenamefont
  {Buttinoni}, \citenamefont {Volpe}, \citenamefont {K\"{u}mmel}, \citenamefont
  {Volpe},\ and\ \citenamefont {Bechinger}}]{Buttinoni2012}%
  \BibitemOpen
  \bibfield  {author} {\bibinfo {author} {\bibfnamefont {I.}~\bibnamefont
  {Buttinoni}}, \bibinfo {author} {\bibfnamefont {G.}~\bibnamefont {Volpe}},
  \bibinfo {author} {\bibfnamefont {F.}~\bibnamefont {K\"{u}mmel}}, \bibinfo
  {author} {\bibfnamefont {G.}~\bibnamefont {Volpe}}, \ and\ \bibinfo {author}
  {\bibfnamefont {C.}~\bibnamefont {Bechinger}},\ }\href@noop {} {\bibfield
  {journal} {\bibinfo  {journal} {J. Phys. Condens. Matter}\ }\textbf {\bibinfo
  {volume} {24}},\ \bibinfo {pages} {284129} (\bibinfo {year}
  {2012})}\BibitemShut {NoStop}%
\bibitem [{\citenamefont {W\"urger}(2015)}]{Wuerger2015}%
  \BibitemOpen
  \bibfield  {author} {\bibinfo {author} {\bibfnamefont {A.}~\bibnamefont
  {W\"urger}},\ }\href@noop {} {\bibfield  {journal} {\bibinfo  {journal}
  {Phys. Rev. Lett.}\ }\textbf {\bibinfo {volume} {115}},\ \bibinfo {pages}
  {188304} (\bibinfo {year} {2015})}\BibitemShut {NoStop}%
\bibitem [{\citenamefont {Trzcinski}\ \emph {et~al.}(1986)\citenamefont
  {Trzcinski}, \citenamefont {Gmelin},\ and\ \citenamefont
  {Queisser}}]{Trzcinski1986}%
  \BibitemOpen
  \bibfield  {author} {\bibinfo {author} {\bibfnamefont {R.}~\bibnamefont
  {Trzcinski}}, \bibinfo {author} {\bibfnamefont {E.}~\bibnamefont {Gmelin}}, \
  and\ \bibinfo {author} {\bibfnamefont {H.~J.}\ \bibnamefont {Queisser}},\
  }\href {\doibase 10.1103/PhysRevLett.56.1086} {\bibfield  {journal} {\bibinfo
   {journal} {Phys. Rev. Lett.}\ }\textbf {\bibinfo {volume} {56}},\ \bibinfo
  {pages} {1086} (\bibinfo {year} {1986})}\BibitemShut {NoStop}%
\end{thebibliography}%

\end{document}